\documentclass{style/nseJournal}

\usepackage{algorithm}
\usepackage{algorithmic}
\usepackage{tabularray}
\usepackage{multirow}
\usepackage{amsmath}
\usepackage[table,xcdraw]{xcolor}

\begin{document}

\title{Mitigating Spatial Error in the iterative-Quasi-Monte Carlo (iQMC) Method for Neutron Transport Simulations with Linear Discontinuous Source Tilting and Effective Scattering and Fission Rate Tallies} 

\addAuthor{\correspondingAuthor{Samuel Pasmann}}{a}
\correspondingEmail{spasmann@nd.edu}
\addAuthor{Ilham Variansyah}{b}
\addAuthor{C.T. Kelley}{c}
\addAuthor{Ryan G. McClarren}{a}

\addAffiliation{a}{University of Notre Dame, \\ Department of Aerospace and Mechanical Engineering \\ Fitzpatrick Hall, Notre Dame, IN 46556}
\addAffiliation{b}{Oregon State University, \\School of Nuclear Science and Engineering \\ 151 Batcheller Hall 1791 SW Campus Way \\ Corvallis, OR 97331}
\addAffiliation{c}{North Carolina State University,\\ Department of
Mathematics\\ 3234 SAS Hall, Box 8205\\ Raleigh NC 27695-8205}

\addKeyword{Neutron Transport}
\addKeyword{Monte Carlo Methods}
\addKeyword{Quasi Monte Carlo}
\addKeyword{Krylov Linear Solvers}
\addKeyword{Source Tilting}

\titlePage

\begin{abstract}

The iterative Quasi-Monte Carlo (iQMC) method is a recently proposed method for multigroup neutron transport simulations. iQMC can be viewed as a hybrid between deterministic iterative techniques, Monte Carlo simulation, and Quasi-Monte Carlo techniques. iQMC holds several algorithmic characteristics that make it desirable for high performance computing environments including a $O(N^{-1})$ convergence scheme, ray tracing transport sweep, and highly parallelizable nature similar to analog Monte Carlo. While there are many potential advantages of using iQMC there are also inherent disadvantages, namely the spatial discretization error introduced from the use of a mesh across the domain. This work introduces two significant modifications to iQMC to help reduce the spatial discretization error. The first is an effective source transport sweep, whereby the source strength is updated on-the-fly via an additional tally. This version of the transport sweep is essentially agnostic to the mesh, material, and geometry. The second is the addition of a history-based linear discontinuous source tilting method. Traditionally, iQMC utilizes a piecewise-constant source in each cell of the mesh. However, through the proposed source tilting technique iQMC can utilize a piecewise-linear source in each cell and reduce spatial error without refining the mesh. Numerical results are presented from the 2D C5G7 and Takeda-1 k-eigenvalue benchmark problems. Results show that the history-based source tilting significantly reduces error in global tallies and the eigenvalue solution in both benchmarks. Through the effective source transport sweep and linear source tilting iQMC was able to converge the eigenvalue from the 2D C5G7 problem to less than $0.04\%$ error on a uniform Cartesian mesh with only $204\times204$ cells.

\end{abstract}

\section{Introduction}

High-fidelity simulation of nuclear reactors and other nuclear systems using general purpose methods is a central goal in computational neutron transport. The computers, both large~\cite{Thomas2020LLNL} and small~\cite{Tramm2023}, that run these problems increasingly rely on graphical processing units (GPUs) for large portions of computation. This has lead to an effort in the neutron transport community to modify existing methods for better performance on GPUs~\cite{GONG2011GPU, Bleile2016Investigation, Hamilton2019Continuous, Hamilton2018Multigroup} or experiment with new methods that may have a more natural fit with GPU architectures~\cite{Willert2012Hybrid, Tramm2021Immortal}. The iterative-Qausi-Monte Carlo (iQMC) method~\cite{Pasmann2023Quasi, Pasmann2023iQMC}, is a recently proposed hybrid-method for multigroup neutron transport simulations that may benefit greatly from implementation on GPUs thanks to its non-divergent algorithmic scheme. However, before porting iQMC to GPUs there are several problem-areas that need to be addressed, primarily spatial discretization error introduced from the use of a 
mesh over the domain of the problem.

iQMC is the combination of deterministic iterative methods, Monte-Carlo (MC) simulation, and quasi-Monte Carlo (QMC) techniques. In iQMC, typical quadrature techniques used in deterministic iterative calculations are replaced with successive Monte-Carlo simulations, referred to as \textit{transport sweeps}. However, by treating the fission and scattering processes as internal fixed sources, the MC transport sweep is reduced to a particle ray-trace and provides a well suited application for QMC. QMC is the replacement of pseudo-random number generators in Monte-Carlo, with low-discrepancy sequences (LDS). LDS use quasi-random or deterministic algorithms to generate sequences with maximum distances between samples. This results in a more efficient sampling of the phase-space and, for $N$ samples, a theoretical $O(N^{-1})$ convergence, compared with the $O(N^{-1/2})$ convergence rate from analog Monte-Carlo~\cite{bickel2009monte}. Additionally, because the LDS provides a very good sample of the phase-space with relatively few particles, a fixed-seed approach can be used in the QMC transport sweep. Meaning, every time a transport sweep is executed, the particles are reset to the same initial position and angle. This allows for the use of more advanced linear Krylov solvers like generalized minimal residual (GMRES) for fixed source problems and the generalized Davidson method for $k$-eigenvalue (criticality) problems. These Krylov solvers have been shown to converge with far fewer transport sweeps in iQMC than the source iteration and power iteration, respectively~\cite{Pasmann2023Reducing, Pasmann2023Quasi, Pasmann2023iQMC}.

Within each QMC transport sweep, particles are emitted with an initial position and direction assigned from samples taken from the LDS. The particle's statistical weight is assigned given the particle's position in a uniform Cartesian mesh with a piecewise-constant source ($Q$). The particles are then traced out of the volume, attenuating the statistical weight with continuous weight absorption, and tallying the cell-averaged scalar flux with a path-length tally estimator. After $N$ particles have been emitted and traced out of the volume, the new scalar flux approximation $\phi^i$ is sent to the iterative method to update the source strength according to the RHS of the neutron transport equation
\begin{equation}
    \label{eq:source}
    Q^i = \frac{\Sigma_s}{2}\phi^i + \chi\nu\Sigma_f\phi^i + q,
\end{equation}
where superscript $i$ denotes the iteration count, $\Sigma_s$ is the macroscopic scattering cross section, $\chi$ is the fission energy distribution, $\nu$ is the average number of neutrons emitted per fission, $\Sigma_f$ is the macroscopic fission cross section, and $q$ is the fixed source. An outline of the QMC transport sweep can be seen in Algorithm~\ref{alg:qmc_sweep}. In short, QMC transport sweeps are iteratively called until a desired tolerance between successive iterations of the source strength or maximum number of iterations has been reached. Compared to history-based Monte-Carlo simulation, iQMC holds several distinct advantages including:
\begin{itemize}
    \item a $O(N^{-1})$ convergence rate, and
    \item a non-divergent algorithm, i.e. the simulation is a ray trace and avoids embedded conditional statements - a promising feature for GPU computations. 
\end{itemize}
iQMC also holds advantages over traditional deterministic methods including:
\begin{itemize}
    \item continuous angular treatment,
    \item a fixed-seed transport sweep that allows for the use of advanced linear Krylov solvers that converge with far fewer iterations than standard iterative schemes like the source and power iteration, and
    \item an ability to easily handle 3D geometries,
    \item a highly parallel nature similar to analog Monte-Carlo.
\end{itemize}

While iQMC holds many advantages there are also challenges. Previous numerical experiments revealed that iQMC is able to converge the scalar flux at the theoretical rate of $O(N^{-1})$ until spatial error, resulting from the use of a piecewise-constant source approximation, begins to dominate. Mitigating the spatial error is most obviously achieved by refining the mesh. However, this also requires a proportional increase in the number of particle histories used in the transport sweep. In large problems, this would quickly lead to escalating increases in the computational cost of the algorithm.

Additionally, the original QMC sweep `order of operations', shown in Algorithm~\ref{alg:qmc_sweep}, necessitates that each cell of the mesh only contains a single material. Meaning, that the material boundaries must align with the mesh cell boundaries, or in the case of curved geometry, results in a piece-wise constant material approximation. This design meant iQMC lost the ability to accurately model arbitrarily complex geometries, a primary advantage of analog Monte Carlo simulations.

To combat these two issues, this work investigates two additions to the iQMC method. The first is a reorganization of the QMC sweep and the addition of an effective-scattering and effective-fission rate tally. This removes the need for the mesh and materials to align and effectively makes the mesh material-agnostic. The second addition is in the form of a history-based linear discontinuous source tilting scheme which should reduce the spatial error without the need to refine the mesh or increase particle count. Section~\ref{sec:effective_sweep} introduces the effective-source sweep modification in iQMC and Section~\ref{sec:source_tilt} formulates the linear discontinuous source tilting tallies. A brief description of implementing iQMC into the Monte Carlo Dynamic Code (MC/DC) is given in Section~\ref{sec:implementation}. Then numerical k-eigenvalue results from the 2D-C5G7 quarter-core reactor problem~\cite{Lewis2001} and Takeda-1 benchmark~\cite{Takeda1991} are presented in Sections~\ref{sec:c5g7_results} and~\ref{sec:takeda1_results} respectively. Conclusions and discussion of future work are presented in Section~\ref{sec:conclusions}.

\section{Methodology}
\label{sec:methods}

\subsection{Flattened Power Iteration}
A recent modification used in this work from previous iQMC for k-eigenvalue problems results~\cite{Pasmann2023iQMC} is in the ``flattening'' of the power iteration. Previously, iQMC completely converged the scattering source, via source iteration or linear Krylov solver (like GMRES), for every update of the eigenvalue and fission source. This nested iterative scheme may be unnecessary as much of the work to converge the scattering source is discarded in the next iteration when the fission source is updated. Recent work, in both the thermal radiative transport field~\cite{Brunner2020} and neutron transport field~\cite{Tramm2017, Gaston2021} have shown that it's reasonable to flatten the power iteration algorithm. In this case, that means updating both the scattering and fission source after only one QMC transport sweep. This version of the iQMC power iteration is depicted in Algorithm~\ref{alg:pi} and was used to generate the results described in Sections~\ref{sec:c5g7_results} and~\ref{sec:takeda1_results}.

\subsection{An Effective Source Transport Sweep}
\label{sec:effective_sweep}

Previously published iQMC results were restricted to simple geometries where the boundaries of a uniform cartesian mesh and the geometry boundaries aligned. This was a result of updating the source strength ($Q$), which requires macroscopic cross section look-ups, \textit{outside} a complete sweep of $N$ particles. Algorithm~\ref{alg:qmc_sweep} shows this original ``flux sweep'' version of the iQMC. With this configuration, modeling complex geometries would require approximations either in the macroscopic cross sections or in the geometry. For example, cross sections from various materials contained in each cell of the mesh could be combined to form one representative set of cross sections for that cell, however, this task would vary by problem and mesh size and has been shown in deterministic methods to introduce non-negligible homogenization error~\cite{kozlowski2011cell}. Another option would be to approximate the geometry by use of a highly refined mesh where each cell is restricted to one set of macroscopic cross sections, resulting in a piecewise-constant material approximation for curved geometry. In this case, refining the mesh to acceptably model curved geometry would require a drastic increase in resolution and a proportional increase in the number of particle histories per iteration. Additionally, no matter how fine the mesh this approximation will always introduce spatial discretization error. Neither of these two solutions is ideal as they introduce errors of one kind or another and generally increase algorithmic complexity. In contrast, an unstructured mesh may be used but as will be discussed in Section~\ref{sec:source_tilt}, this would significantly complicate the proposed history-based source tilting method and require that the mesh become problem-dependent. Whereas with the current formulation, little attention needs to be paid to the mesh other than ensuring a fine enough resolution for convergence.

\begin{figure}
  \centering
  \includegraphics[width=0.9\textwidth]{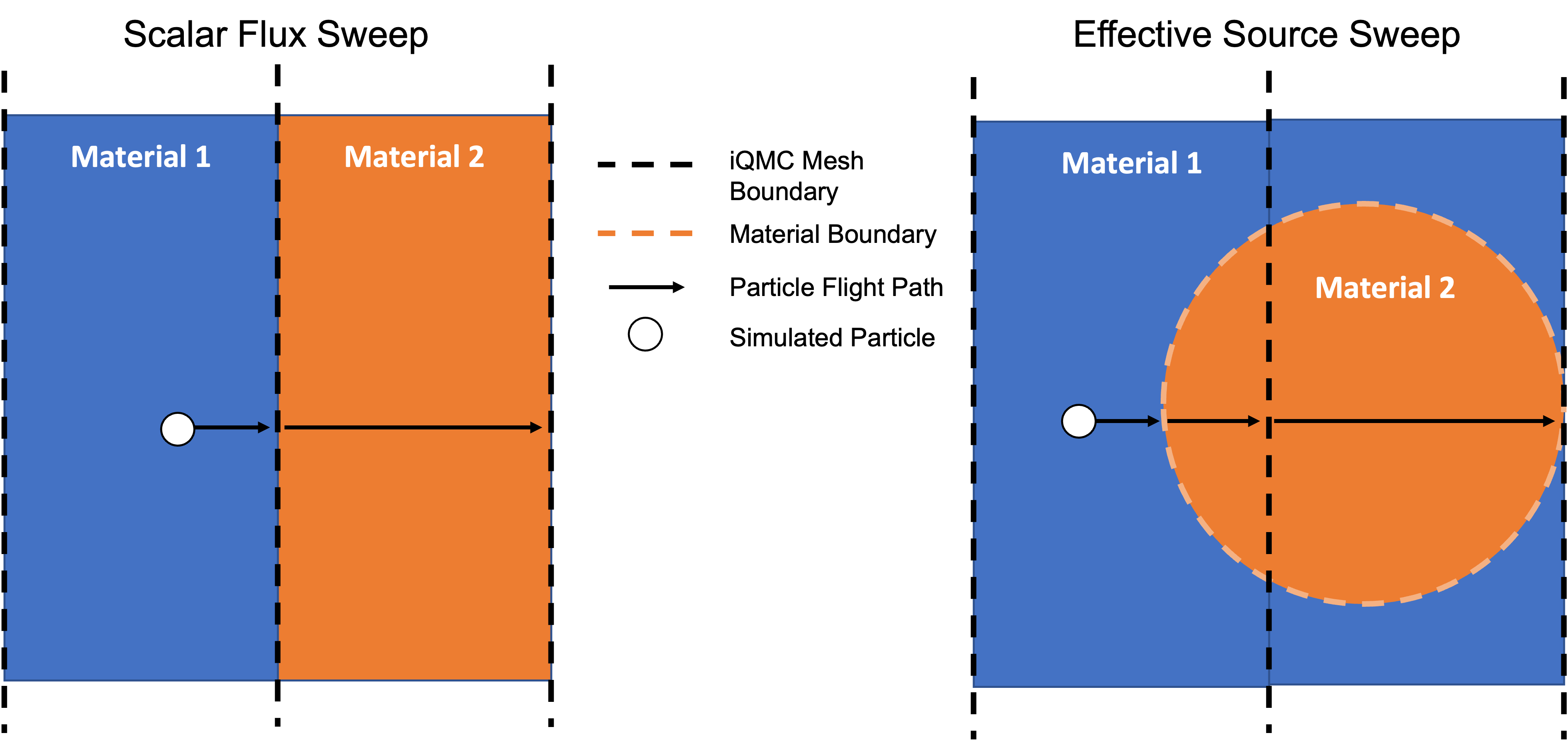}
  \caption{The initial implementation of iQMC transport sweep required that the Cartesian mesh and material boundaries be aligned. The effective source sweep decouples the mesh and material and the particle can now tally flux contributions from an arbitrary number of materials into the same mesh cell.}
  \label{fig:particle_tracking}
\end{figure}

Instead, we explore a reformulation of the iQMC transport sweep which allows for the mesh and materials contained within to be independent. Rather than updating the source strength with the scalar flux after a complete sweep of $N$ particles, the source strength is updated on-the-fly via an effective scattering and fission rate tally. Algorithm~\ref{alg:effective_sweep} shows this ``effective source sweep'', in which line $3$ of Algorithm~\ref{alg:qmc_sweep} is moved inside the particle loop to line $8$ of Algorithm~\ref{alg:effective_sweep}. Figure~\ref{fig:particle_tracking} shows a simplified depiction of the particle tracking between the two sweep methods.

\begin{minipage}{0.66\textwidth}
\begin{algorithm}[H]
    \centering
    \caption{Scalar Flux Sweep ($\phi_\textrm{in}$)}\label{alg:qmc_sweep}
    \begin{algorithmic}[1]
        \FOR {Cells in mesh}
            \STATE prepare source strength $Q$  
            using $\phi_\textrm{in}$ and Eq.~\ref{eq:source}
        \ENDFOR
		\FOR {$N$ particles}
			\STATE Assign position and angle based on the LDS
			\STATE Initialize weight based on $Q$
			\WHILE {Particle = alive}
				\STATE Move particle across the zone
				\STATE Tally scalar flux (Eq.~\ref{eq:flux_tally})
				\STATE Attenuate particle weight
                \IF {Particle exits volume}
                    \STATE Particle = dead
                \ENDIF
			\ENDWHILE
		\ENDFOR
		\STATE Return: $\phi_\textrm{out}$
    \end{algorithmic}
\end{algorithm}
\end{minipage}

\begin{minipage}{0.66\textwidth}
\begin{algorithm}[H]
    \centering
    \caption{Effective Source Sweep ($Q_\textrm{in}$)}\label{alg:effective_sweep}
    \begin{algorithmic}[1]
		\FOR {$N$ particles}
			\STATE Assign position and angle based on the LDS
			\STATE Initialize weight based on $Q$
			\WHILE {Particle is alive}
				\STATE Move particle across the zone
				\STATE Tally scalar flux (Eq.~\ref{eq:flux_tally})
                \STATE Tally effective source (Eq.~\ref{eq:source})
				\STATE Attenuate particle weight
                \IF {Particle exits volume}
                    \STATE Particle = dead
                \ENDIF
			\ENDWHILE
		\ENDFOR
		\STATE Return: $Q_\textrm{out}$, $\phi_\textrm{out}$
    \end{algorithmic}
\end{algorithm}
\end{minipage}

\begin{minipage}{0.66\textwidth}
\begin{algorithm}[H]
    \centering
    \caption{Flattened iQMC Power Iteration}\label{alg:pi}
    \begin{algorithmic}[1]
		\STATE Initialize Low-Discrepancy-Sequence (LDS)
		\WHILE {$k_\text{eff}$ and Scalar Flux \textbf{not} Converged}
            \STATE QMC Transport Sweep (Alg.~\ref{alg:effective_sweep})
            \STATE Update $k_\text{eff}$
            \STATE Calculate relative difference in $k_\text{eff}$ and scalar flux
		\ENDWHILE
		\STATE Return: $\phi_\textrm{out}$
    \end{algorithmic}
\end{algorithm}
\end{minipage}

\subsection{Linear Source Tilting}
\label{sec:source_tilt}
\begin{figure}
  \centering
  \includegraphics[width=\textwidth]{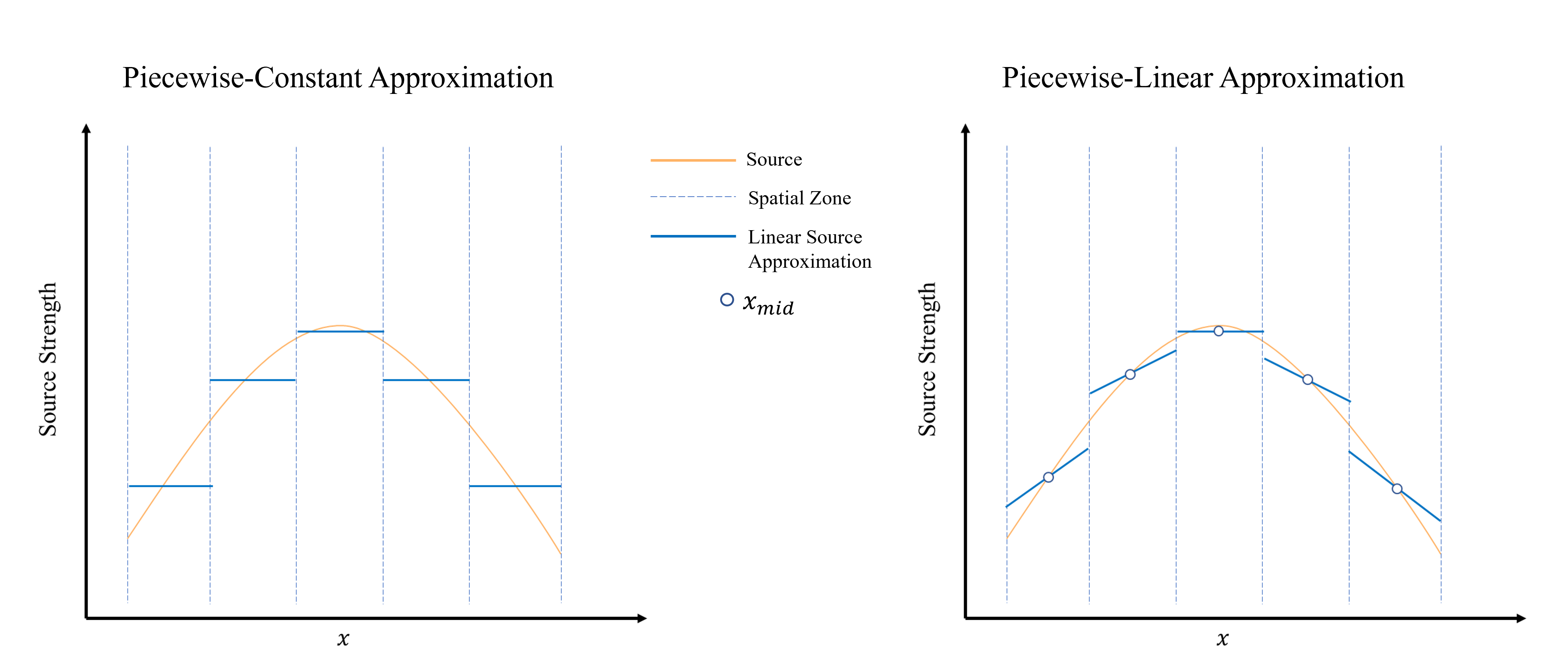}
  \caption{A cartoon depiction of a piecewise-constant and piecewise-linear approximation of some theoretical source.}
  \label{fig:source_tilting}
\end{figure}

This section focuses on the development and evaluation of a history-based linear discontinuous source tilting method. The term ``source tilting'' refers to a host of commonly used techniques in implicit Monte Carlo (IMC) for thermal radiative transfer problems~\cite{irvine2016reducing, Wollaeger2016, smedley2015asymptotic, shi2020continuous}. Similar to iQMC, IMC utilizes a mesh across the problem to discretize material-energy. Using a piecewise-constant material-energy evaluation can lead to non-physical heating where particles may travel faster than would be physically allowed, this is referred to as ``teleportation error''~\cite{wollaber2016four}. Source tilting mitigates teleportation error by creating a piecewise-linear (or higher order) subprofile in each cell of the grid to sample particles from. A similar phenomenon exists in iQMC where the source strength is iterated on and stored within a piecewise constant mesh. On a coarse mesh or in problems with rapid changes in the scalar flux, this piecewise-constant approximation may be inadequate. By constructing a higher order source profile in each cell we hope to reduce the spatial error observed in previous iQMC results~\cite{Pasmann2023Quasi}. Figure~\ref{fig:source_tilting} provides a simple depiction of a piecewise-constant and piecewise-linear scheme with a 1-dimensional example.

IMC source tilting techniques typically use interpolation to estimate the higher order terms in each cell. In this work, we investigate a history-based source tilting scheme to directly estimate the linear slope. Simple 2-dimensional results indicated the method is effective in reducing the spatial discretization error, however, it does not eliminate it~\cite{Pasmann2023Reducing} and has yet to to be tested on more complex or 3D problems. Section~\ref{sec:piecewise-constant} formulates the base piecewise-constant scheme used in iQMC then Section~\ref{sec:piecewise-linear} introduces the tallies necessary to linearly tilt the source in 3-dimensions. Finally Section~\ref{sec:takeda1_results} evaluates the method on the 3D Takeda-1 benchmark problem~\cite{Takeda1991}.

\subsubsection{Piecewise-Constant Scheme}
\label{sec:piecewise-constant}
iQMC uses a continuous absorption scheme to attenuate the particle's statistical weight with every step,
\begin{equation}
    \label{eq:cwr}
    w_\text{new} = w_\text{old}\text{e}^{-\Sigma_t l}.
\end{equation}
Here $w_\text{old}$ is the initial particle weight, $l$ is the path-length, and $\Sigma_t$ is the total cross section. We integrate Equation~\ref{eq:cwr} and arrive at our path-length scalar flux tally estimator,
\begin{equation}
    \label{eq:flux_tally}
    \Delta\phi_{i,j,k} = \frac{1}{\Delta x \, \Delta y \, \Delta z}\int_{0}^{l}w_{\textrm{old}}e^{-\Sigma_{t}l^\prime}dl^\prime = \frac{w_{\textrm{old}}}{V}\left(\frac{1-e^{-\Sigma_{t}l}}{\Sigma_{t}}\right) .
\end{equation}
Here $i$, $j$, and $k$ denote the spatial indices of the current cell and $\Delta x = x_{i+\frac{1}{2}} - x_{i-\frac{1}{2}}$, $\Delta y = y_{j+\frac{1}{2}} - y_{j-\frac{1}{2}}$, and $\Delta z = z_{k+\frac{1}{2}} - z_{k-\frac{1}{2}}$ are the cell widths for the $x$, $y$, and $z$ dimensions respectively. Equation~\ref{eq:flux_tally} from the effective source sweep of Section~\ref{sec:effective_sweep}, is then used to tally the effective source strength,
\begin{equation}
    \label{eq:effective_source_tally}
    \Delta Q_{i,j,k} = \frac{\Sigma_s}{2}\Delta\phi_{i,j,k} + \chi\nu\Sigma_f\Delta\phi_{i,j,k} + q .
\end{equation}
After $N$ particle sweeps, in a piecewise-constant approximation, the scalar flux $\phi$ and source $Q$ in each cell is entirely described by the tally results of Equation~\ref{eq:flux_tally} and~\ref{eq:effective_source_tally} respectively. To simplify our piecewise-constant representation of the scalar flux, let
\begin{equation}
    \label{eq:A}
    \phi_{\mathrm{Constant}} = A_{i,j,k}.
\end{equation}

\subsubsection{Piecewise-Linear Scheme}
\label{sec:piecewise-linear}
To ``tilt'' the source for a piecewise-linear representation we will develop a flux tilting tally and pass the results through the same effective source tally in Equation~\ref{eq:effective_source_tally}. For a full 3D representation of the discontinuous linear scheme let $\Delta x_\text{mid} = (x_\text{new}-x_{\mathrm{mid},i})$, $\Delta y_\text{mid} = (y_\text{new}-y_{\mathrm{mid},j})$, and $\Delta z_\text{mid} = (z_\text{new}-z_{\mathrm{mid},k})$ represent the particle's distance from the cell midpoint in the $x$, $y$, and $z$ dimensions after moving distance $l$. Therefore $x_\text{new} = (x_\text{old} + \mu_x l)$, $y_\text{new} = (y_\text{old} + \mu_y l)$, and $z_\text{new} = (z_\text{old} + \mu_z l)$, where $\mu_x$, $\mu_y$, and $\mu_z$ denote the axial components of particle flight direction unit vector. The source may now be represented as
\begin{equation}
    \label{eq:linear_scheme}
        \phi_{\mathrm{Linear}} = A_{i,j,k} 
        + B_{i,j,k} \Delta x_\text{mid} 
        + C_{i,j,k} \Delta y_\text{mid} 
        + D_{i,j,k} \Delta z_\text{mid} 
\end{equation}
The $A_{i,j,k}$ term is the same as the piecewise-constant scheme in Equation~\ref{eq:A}. The $B_{i,j,k} \,$, $C_{i,j,k} \,$, $D_{i,j,k} \,$ terms represent the linear source-tilt in the $x$, $y$, and $z$ directions respectively, and can be similarly tallied with the analytic result of
\begin{equation} 
    \label{eq:B}
    B_{i,j,k} = \frac{12}{\Delta x_{i}^3 \Delta y_{j} \, \Delta z_{k}} \int_{0}^{l} 
    \Delta x_\text{mid}  w_{\text{old}}e^{-\Sigma_{t}l^\prime}dl^\prime ,
\end{equation}
\begin{equation}
    \label{eq:C}
    C_{i,j,k} = \frac{12}{\Delta x_{i} \Delta y_{j}^3 \, \Delta z_{k}} \int_{0}^{l} 
    \Delta y_\text{mid}  w_{\text{old}}e^{-\Sigma_{t}l^\prime}dl^\prime ,
\end{equation}
and
\begin{equation} 
    \label{eq:D}
    D_{i,j,k} = \frac{12}{\Delta x_{i} \Delta y_{j} \, \Delta z_{k}^3} \int_{0}^{l} 
    \Delta z_\text{mid}  w_{\text{old}}e^{-\Sigma_{t}l^\prime}dl^\prime . 
\end{equation}

Equations~\ref{eq:B}, ~\ref{eq:C}, and~\ref{eq:D} essentially represent a first-order moment tallies in space. Using this technique in an unstructured mesh would present a significant challenge as the centroid of each cell is needed to calculate the source tilting tally. In contrast, the use of a Cartesian mesh means source titling terms may be calculated in the same manner for all cells.

With each particle step, the result of the flux tilt tally must also be passed through the effective source tally in Equation~\ref{eq:effective_source_tally} similar to the piecewise-constant term. Therefore, tilting in one dimension effectively adds two tallies to the transport sweep. In a piecewise-constant simulation iQMC has a minimum of two tallies: scalar flux and effective source, while the piecewise linear scheme can add up to 6 additional tallies (two for each dimension), possibly incurring significant computational cost. An additional consideration is that by increasing the number of terms used to describe the source strength we are also increasing the number of terms that need to be converged in the iterative solver. If in a piecewise-constant simulation the size of the source iteration or Krylov vector is $N_G \times N_c$ where $N_G$ is the number of energy groups and $N_c$ is the number of spatial cells in the mesh, then adding source tilting increases this vector to $N_G \times N_c \times N_Q$ where $N_Q$ is the number of terms used to describe the source strength $Q$. With linear Krylov solvers, this could pose significant memory constraints on large problems as we need to store a number of these large vectors.

\section{Numerical Results}
\label{sec:results}

\subsection{Implementation}
\label{sec:implementation}

iQMC was implemented in the Monte Carlo Dynamic Code (MC/DC), a Python-based exploratory Monte Carlo neutron transport code in development in the Center for Exascale Monte-Carlo for Neutron Transport (CEMeNT)~\cite{variansyah2023MCDC}. MC/DC leverages Numba, a just-in-time compiler for scientific computing in Python~\cite{lam2015numba}, to greatly enhance Python performance. MC/DC also offers massive scalability with domain replication via MPI4Py~\cite{dalcin2005mpi} and backend portability by leveraging Python code generation libraries and kernel abstraction scheme \cite{variansyah2023development}.

MC/DC was primarily designed as a rapid prototype environment for Monte-Carlo neutron transport algorithms and variance reduction techniques. Nonetheless, MC/DC has shown good performance with 56-212x speedups over pure Python on a variety of problems and 2.5x slower, but with similar parallel scaling, than the C++ based Shift Mont Carlo code~\cite{pandya2016implementation} for simple problems~\cite{variansyah2023development}.

\subsection{2D-C5G7 k-Eigenvalue}
\label{sec:c5g7_results}
\begin{figure}
  \centering
  \includegraphics[width=0.8\textwidth]{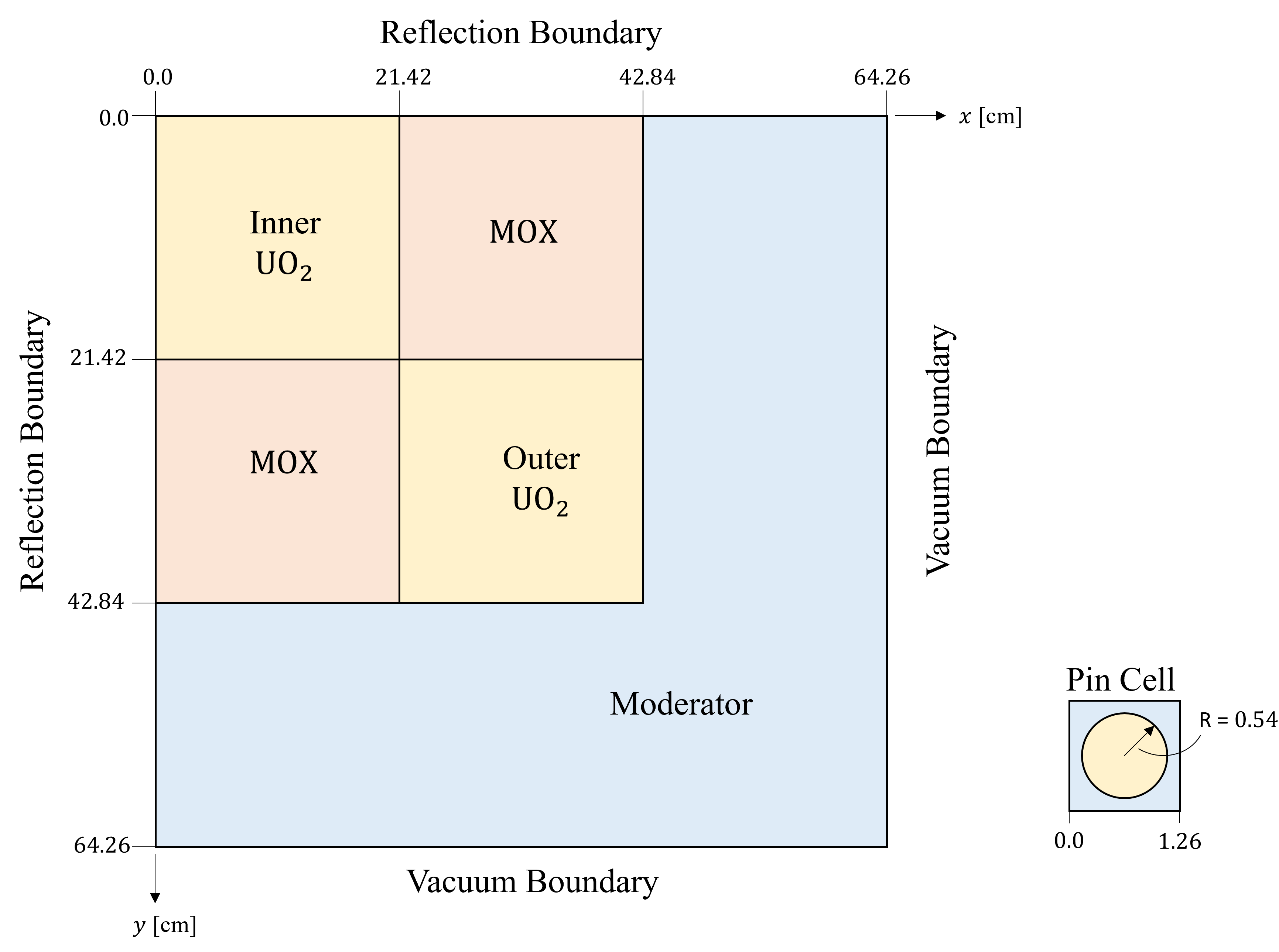}
  \caption{2D C5G7 quarter-core reactor problem. Each assembly consists of a $17\times17$ pin cell array.}
  \label{fig:c5g7_geometry}
\end{figure}

\begin{figure}
  \centering
  \includegraphics[width=\textwidth]{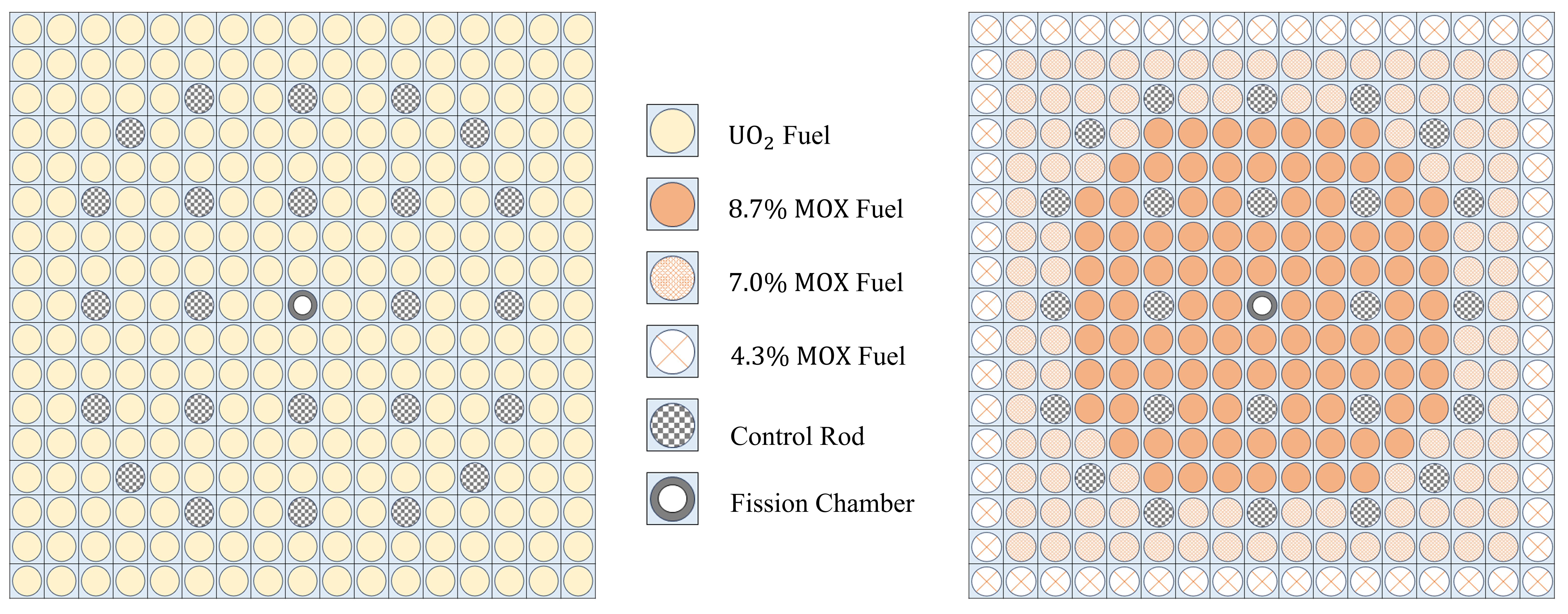}
  \caption{$\text{UO}_\text{2}$ (left) and MOX (right) assembly configuration for 2D C5G7 problem.}
  \label{fig:c5g7_assemblies}
\end{figure}

To evaluate the effective source sweep and linear source tilting in iQMC we present numerical results from the 2D C5G7 benchmark problem, a 7-group quarter-core reactor problem~\cite{Lewis2001}. The problem consists of four assemblies, each made up of a $17\times17$ pin cell array of various fuels and control rods. Figure~\ref{fig:c5g7_geometry} depicts assembly and moderator configuration while Figure~\ref{fig:c5g7_assemblies} depicts the two assembly configurations, $\text{UO}_\text{2}$ and mixed-oxide (MOX), in more detail. iQMC eigenvalue and assembly-wise fission power results were compared to published benchmark results from a mutligroup Monte Carlo (MGMC) results generated with 300 million particle histories with MCNP~\cite{smith2004benchmark}.

iQMC was run with four uniform Cartesian mesh sizes: $51\times51$, $102\times102$, $204\times204$, and finally $408\times408$. Then, each mesh size was run with both the piecewise-constant and piecewise-linear iQMC approximations. The coarse $51\times51$ mesh consists of only $23,409$ cells which corresponds to exactly one mesh cell per pin. The ``fine'' mesh is equivalent to $8\times8$ cells per pin. For comparison, deterministic literature states that $14\times14$ cells are needed to acceptably model a fuel pin~\cite{van20072}. This equates to $509,796$ cells, or over 3x as many cells as in the currently defined ``fine'' mesh. To fairly compare the effect of changing the mesh size, each simulation was run with a constant ratio of particle histories ($N$) to mesh cells ($N_c$) such that $N/N_c=200$, so results generated with a finer mesh also contain more particle histories. This equates to $N=4,681,800$ and $N=33,292,800$ particles per iteration for the coarse and fine mesh respectively. All used the same criteria for convergence $\Delta k_\text{eff} / k_\text{eff} \leq 10^{-5}$ and $\Delta \phi / \phi \leq 10^{-5}$.

\begin{figure}
  \centering
  \includegraphics[width=\textwidth]{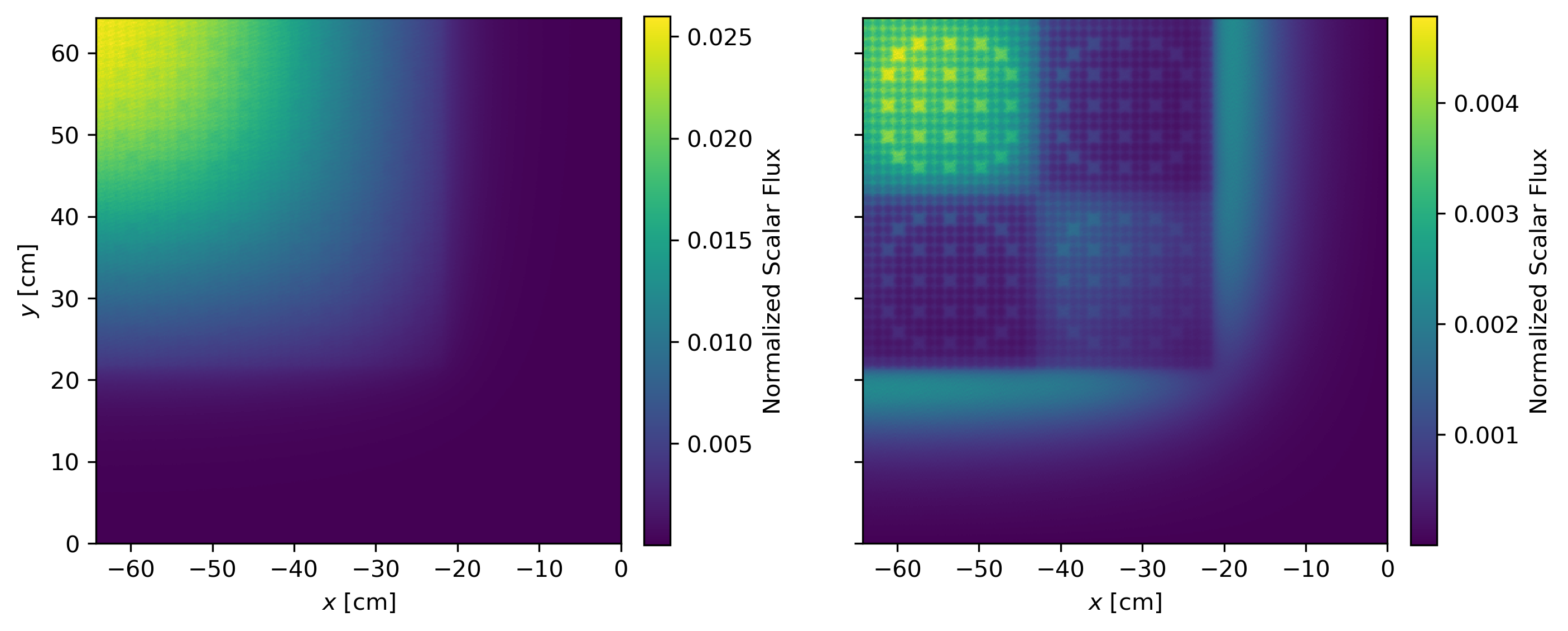}
  \caption{iQMC scalar flux results with the fine mesh, separated by fast groups 1-4 (left) and thermal groups 5-7 (right).}
  \label{fig:c5g7_flux}
\end{figure}

Figure~\ref{fig:c5g7_flux} shows the scalar flux output of the fast and thermal flux groups from the iQMC fine mesh simulation with linear source tilting. Table~\ref{table:c5g7_eig} shows the eigenvalue solutions and percent error relative to the MCNP solution. The coarse mesh's constant and linear results are relatively inaccurate with error $\geq 1\%$. The remaining results follow a consistent pattern where refining the mesh yields more accurate results and subsequently using a piecewise-linear source tilt improves the results further. In all cases except the coarse mesh, the linear results from any given mesh are more accurate than the subsequently refined mesh's constant approximation.

\begin{figure}
  \centering
  \includegraphics[width=\textwidth]{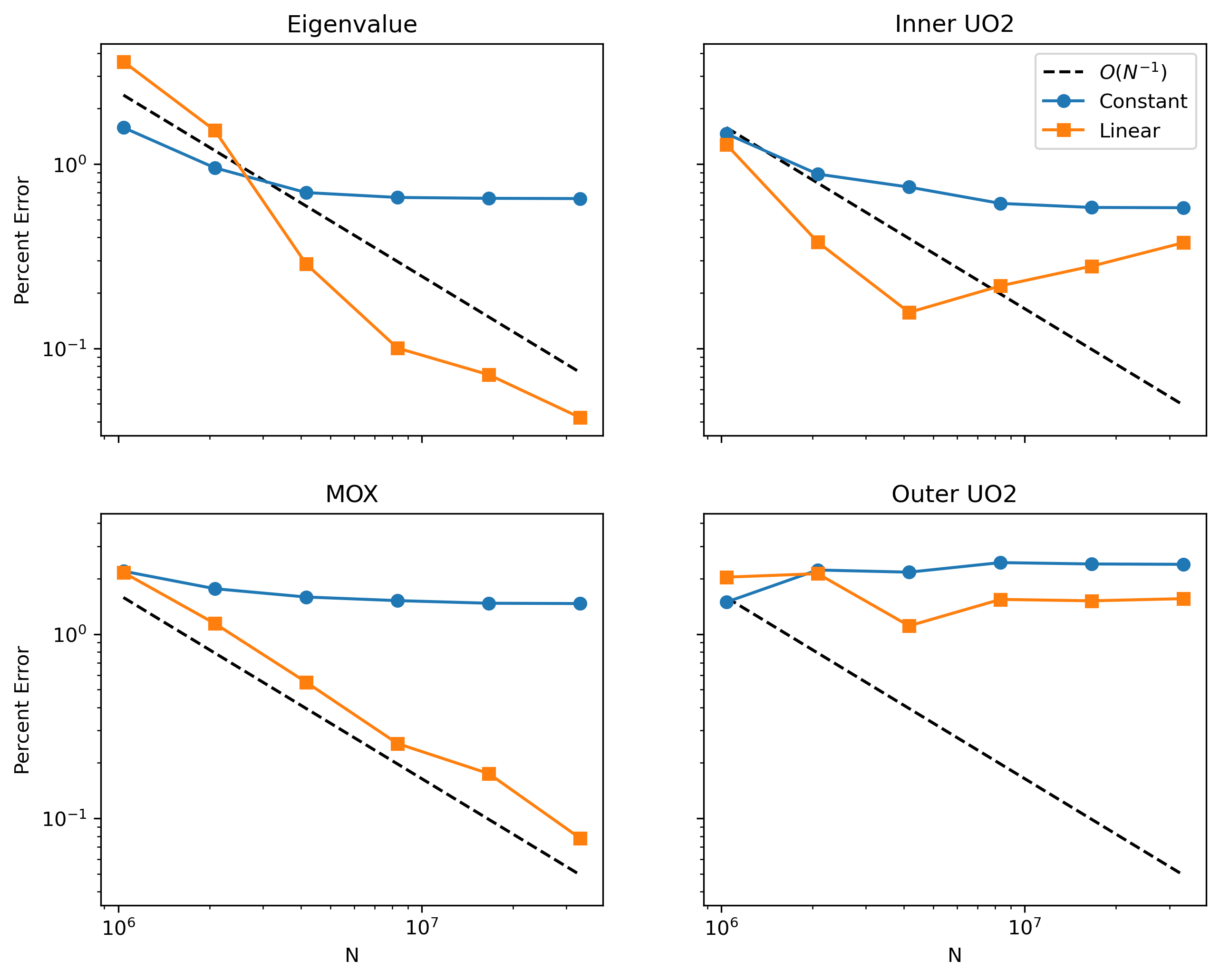}
  \caption{Eigenvalue and assembly power percent error as a function of the number of particle histories per iteration $N$ on the $204\times204$ mesh.}
  \label{fig:c5g7_conv}
\end{figure}

Table~\ref{table:c5g7_fission} shows the assembly fission powers and percent error relative to the MCNP results. The MOX results are an average of the two MOX assemblies. The Inner $\text{UO}_2$ and MOX fission power show the same patterns established in the eigenvalue results. However, the Outer $\text{UO}_2$ assembly appears to stall in error convergence. Figure~\ref{fig:c5g7_conv} displays the percent error as a function of particle histories on the $204\times204$ mesh, for the eigenvalue solution and assembly powers. Both the eigenvalue and MOX assembly solutions converge at the theoretical QMC rate of $O(N^{-1})$. However, both $\text{UO}_2$ assemblies stall in convergence indicating this mesh is too coarse to adequately mitigate the spatial error in these regions.

 Together, these results show that by refining the mesh in iQMC the spatial error is reduced and k-eigenvalue and assembly power solutions are more accurate. This phenomenon is expected, as is similar to other deterministic methods. However, by using the effective source sweep iQMC can accurately model the problem with far fewer cells in the mesh than with the previous scalar flux sweep, also noting the fine mesh is still far too coarse for most deterministic methods. Further, using the linear discontinuous source tilting method the iQMC results can be made significantly more accurate, in some cases outperforming mesh refinement.

\begin{table}
\centering
\caption{2D C5G7 Eigenvalue Solutions}
\begin{tabular}{lllrl}
\hline
 Method   & Mesh    & Source   & Eigenvalue & \% Error   \\
\hline
 MGMC   & ---     & ---      &    1.18655 & ± 0.008   \\
 iQMC   & 51x51   & Constant &    1.20074 &   1.196   \\
 iQMC   & 51x51   & Linear   &    1.2062  &   1.656   \\
 iQMC   & 102x102 & Constant &    1.2015  &    1.26   \\
 iQMC   & 102x102 & Linear   &    1.19205 &   0.463   \\
 iQMC   & 204x204 & Constant &    1.19438 &   0.659   \\
 iQMC   & 204x204 & Linear   &    1.18774 &     0.1   \\
 iQMC   & 408x408 & Constant &    1.18864 &   0.176   \\
 iQMC   & 408x408 & Linear   &    1.18595 &    0.05   \\
\hline
\end{tabular}
\label{table:c5g7_eig}
\end{table}

\vspace{0.25in}

\begin{table}
\raggedright
\caption{Assembly Power Percent Error}
\begin{tabular}{lllrlrlrl}
\hline
 Method   & Mesh    & Source  & Inner $\text{UO}_2$ & \% Error   & MOX & \% Error   & Outer $\text{UO}_2$ & \% Error   \\
\hline
 MGMC   & ---     & ---      &      492.8  & ± 0.1     & 211.7  & ± 0.18 &       139.8  & ± 0.2     \\
 iQMC   & 51x51   & Constant &      490.04 & 0.56      & 208.26 & 1.62   &       149.43 & 6.89      \\
 iQMC   & 51x51   & Linear   &      509    & 3.29      & 201.82 & 4.67   &       143.36 & 2.55      \\
 iQMC   & 102x102 & Constant &      501.38 & 1.74      & 204.55 & 3.38   &       145.52 & 4.09      \\
 iQMC   & 102x102 & Linear   &      494.61 & 0.37      & 209.89 & 0.86   &       141.62 & 1.3       \\
 iQMC   & 204x204 & Constant &      495.82 & 0.61      & 208.48 & 1.52   &       143.21 & 2.44      \\
 iQMC   & 204x204 & Linear   &      491.72 & 0.22      & 211.16 & 0.25   &       141.96 & 1.54      \\
 iQMC   & 408x408 & Constant &      491.37 & 0.29      & 211.09 & 0.29   &       142.45 & 1.89      \\
 iQMC   & 408x408 & Linear   &      489.88 & 0.59      & 212.01 & 0.15   &       142.09 & 1.64      \\
\hline
\end{tabular}
\label{table:c5g7_fission}
\end{table}

\subsection{Takeda-1 k-Eigenvalue}
\label{sec:takeda1_results}
\begin{figure}
  \centering
  \includegraphics[width=\textwidth]{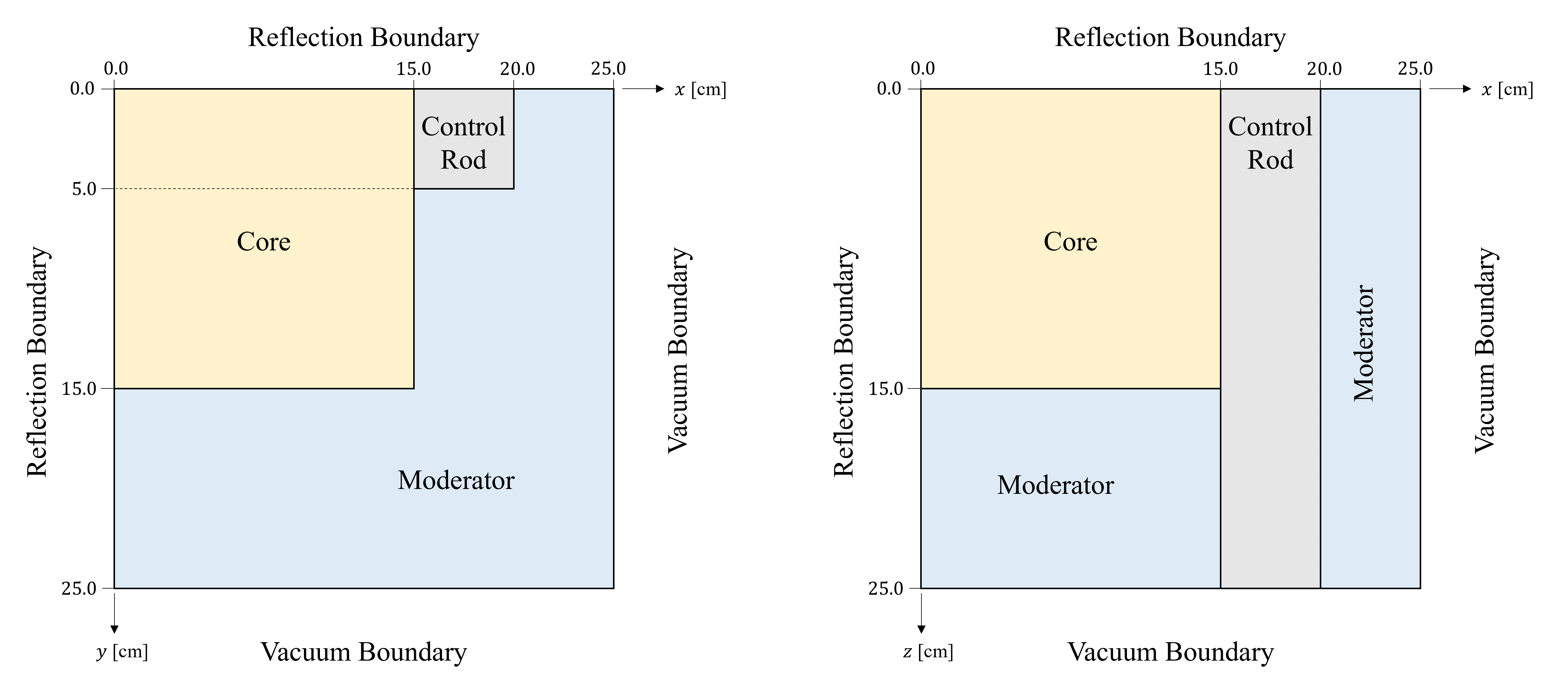}
  \caption{Takeda-1 Benchmark geometry specifications.}
  \label{fig:takeda1_geometry}
\end{figure}

Figure~\ref{fig:takeda1_geometry} depicts the Takeda-1 benchmark a simple, 3D, 2-group, k-eigenvalue reactor problem~\cite{Takeda1991} featuring a core, control rod, and moderator regions. There are two variations of the problem, one where the control rod is removed and the region is void, and the other where the control rod is inserted the region is highly absorbing. The results presented were run with a control rod inserted. iQMC and analog Monte Carlo reference results were generated on a $25\times25\times25$ spatial grid. The reference solution was generated with analog multigroup Monte Carlo (MGMC) in MC/DC with 50 million particle histories per batch with 10 inactive and 20 active batches. iQMC results were converged to a tolerance of $\Delta k_\text{eff} / k_\text{eff} \leq 10^{-5}$ and $\Delta \phi / \phi \leq 10^{-5}$.

\begin{figure}
  \centering
  \includegraphics[width=\textwidth]{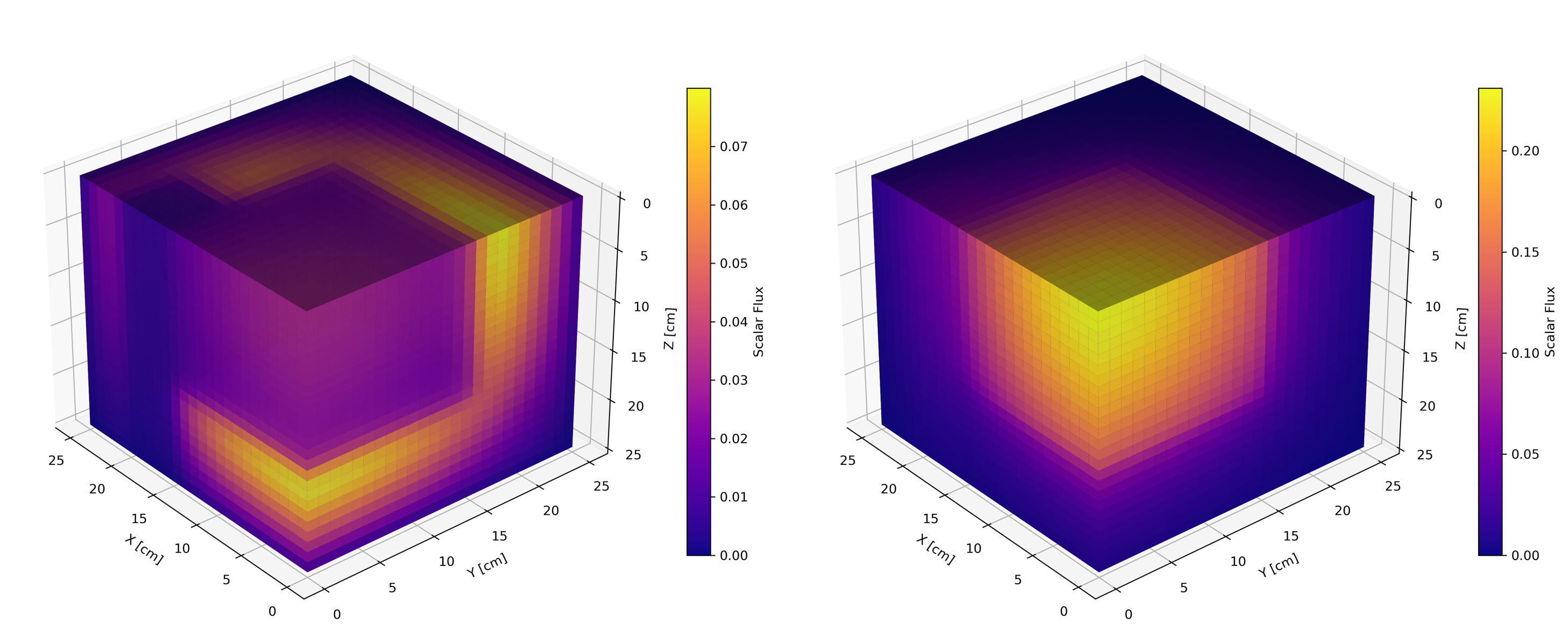}
  \caption{Fast scalar flux (left) and thermal flux (right) from iQMC simulation of the Takeda-1 problem with linear source tilting.}
  \label{fig:takeda1_flux}
\end{figure}

For this experiment, iQMC was run with a low $N=3.125\times10^6$ and a high $N=3.125\times10^7$ number of particle histories per iteration. These values were chosen to maintain a ratio of particle histories to spatial cells $N/N_c=200$ and $N/N_c=2000$ for the low and high particle count tests respectively. Both the low and high particle count tests were run with the standard piecewise-constant approximation and the piecewise-linear source tilt. Figure~\ref{fig:takeda1_flux} shows the scalar flux solution from the iQMC high particle count test with linear source tilting.

\begin{figure}
  \centering
  \includegraphics[width=\textwidth]{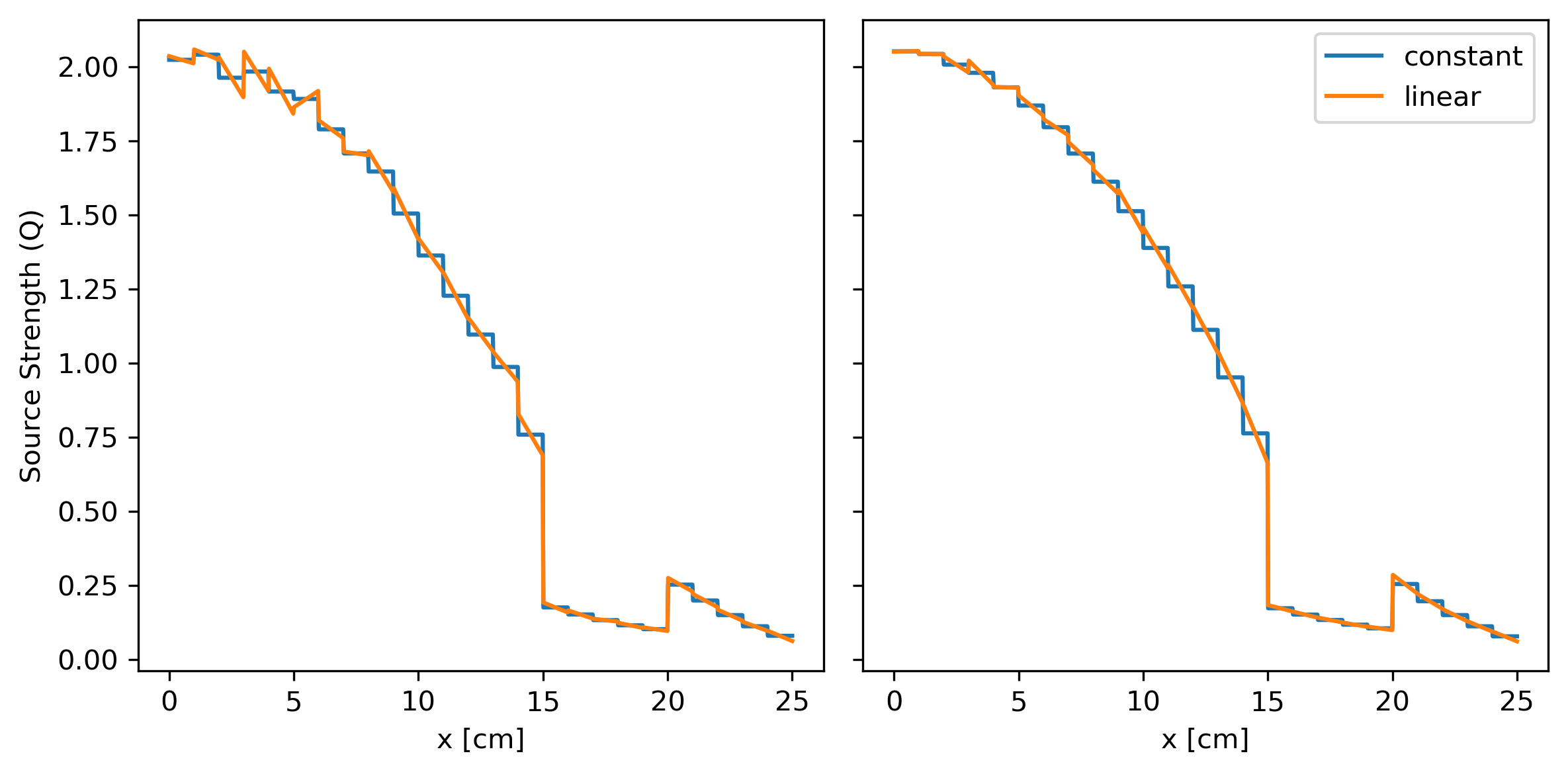}
  \caption{Final fast-group source strength ($Q$), constant and x-tilted, results between $x\in [0,25]$, $y\in [0.0, 1.0]$, $z\in [0.0, 1.0]$ from the Takeda-1 benchmark with low particle count (left) and high particle count (right). }
  \label{fig:takeda1_fast_source_tilt}
\end{figure}

\begin{figure}
  \centering
  \includegraphics[width=\textwidth]{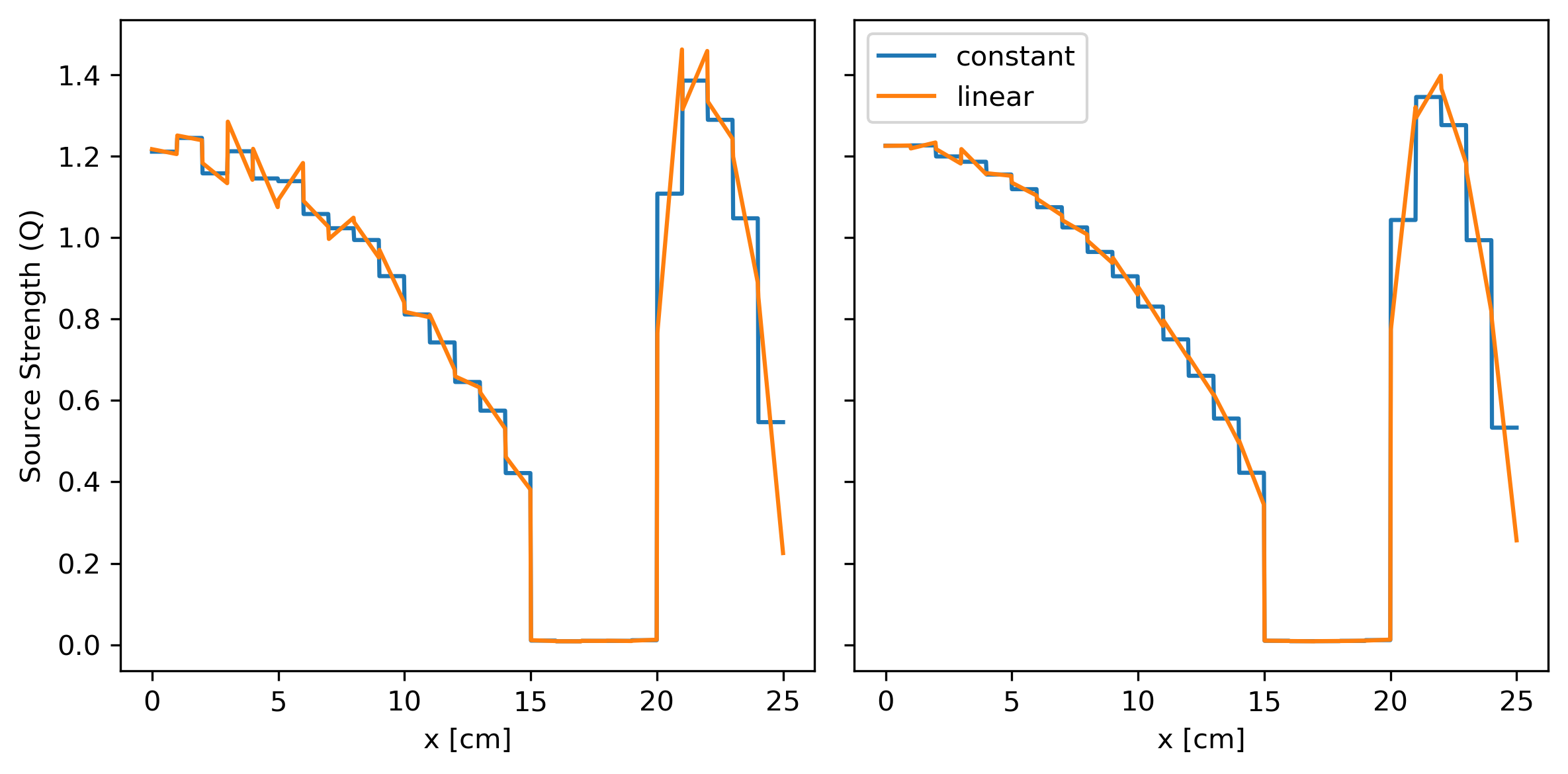}
  \caption{Final slow-group source strength ($Q$), constant and x-tilted, results between $x\in [0,25]$, $y\in [0.0, 1.0]$, $z\in [0.0, 1.0]$ from the Takeda-1 benchmark with low particle count (left) and high particle count (right). }
  \label{fig:takeda1_slow_source_tilt}
\end{figure}

Table~\ref{table:takeda1_results} shows the region averaged scalar flux and percent error results from iQMC and the Monte Carlo reference solution. As is clear from these results, the source tilting method significantly reduced error in every region of the problem for both the low and high particle runs. Of course, the addition of source tilting may incur significant computational cost. To explore this trade-off, we define a figure of merit (FOM) to be 
\begin{equation}
    \text{FOM} = 1000 / (\epsilon\times t),
\end{equation}
where $\epsilon$ is the flux percent error from Table~\ref{table:takeda1_results} and $t$ is the total simulation time. Consistent with relative error results, Table~\ref{table:takeda1_fom} shows an increasing FOM for both piecewise-linear results compared to their piecewise-constant counterparts. With a low particle count the FOM increases by a factor of about 2, while the FOM from the high particle count test increases by a factor of about 20. Why source tilting has a higher FOM in the high particle count test is made clear in Figure~\ref{fig:takeda1_fast_source_tilt} and~\ref{fig:takeda1_slow_source_tilt}, which depicts the converged source for a 1-dimensional slice of the problem. For a low number of particle histories, the source tilting tallies can contribute significant noise to the solution. Increasing the particle count further resolves the linear terms and reduces noise.

\begin{table}
\centering
\caption{Region averaged scalar flux from Takeda-1 benchmark, integrated over energy.}
\begin{tabular}{lllrlrlrl}
\hline
 Method   & N    & Source   &    Core &   \% Error &      CR &   \% Error &     Mod &   \% Error \\
\hline
 MGMC     & ---  & ---      & 0.15383 &    ±0.0179 & 0.03922 &    ±0.0317 & 0.14601 &    ±0.0387 \\
 iQMC     & Low  & Constant & 0.1572  &    2.19   & 0.04084 &    4.13   & 0.14239 &    2.48   \\
 iQMC     & Low  & Linear   & 0.1527  &    0.74   & 0.03903 &    0.48   & 0.14704 &    0.7    \\
 iQMC     & High & Constant & 0.15742 &    2.33   & 0.04087 &    4.21   & 0.14219 &    2.62   \\
 iQMC     & High & Linear   & 0.15397 &    0.09   & 0.0393  &    0.2    & 0.14589 &    0.09   \\
\hline
\end{tabular}
\label{table:takeda1_results}
\end{table}

\vspace{0.25in}

\begin{table}
\centering
\caption{iQMC region averaged flux figures of merit from the Takeda-1 benchmark.}
\begin{tabular}{llrrr}
\hline
 N    & Source   &    Core &     CR &     Mod \\
\hline
 Low  & Constant &  1.6902 & 0.8953 &  1.4884 \\
 Low  & Linear   &  3.9198 & 6.042  &  4.1362 \\
 High & Constant &  1.1754 & 0.6501 &  1.0454 \\
 High & Linear   & 20.4172 & 9.2025 & 21.6004 \\
\hline
\end{tabular}
\label{table:takeda1_fom}
\end{table}

\vspace{0.25in}

\begin{table}
\centering
\caption{Takeda-1 Eigenvalue Solutions}
\begin{tabular}{lllrl}
\hline
 Method   & N    & Source   &   Eigenvalue & \% Error   \\
\hline
 MGMC     & ---  & ---      &     0.962369 & ± 0.0     \\
 iQMC     & Low  & Constant &     0.960383 & 0.206     \\
 iQMC     & Low  & Linear   &     0.966915 & 0.472     \\
 iQMC     & High & Constant &     0.959657 & 0.282     \\
 iQMC     & High & Linear   &     0.962156 & 0.022     \\
\hline
\end{tabular}
\label{table:takeda1_eig}
\end{table}


\section{Conclusions}
\label{sec:conclusions}

This paper introduced several new methods to the iterative Quasi-Monte Carlo (iQMC) method for multigroup neutron transport simulations. First, a flattening of the power iteration wherein the scattering and fission sources are both updated after only one transport sweep. Then, two methods were developed to reduce the spatial discretization error introduced by the use of a Cartesian mesh. The first is the use of an effective source transport sweep which adds an additional tally so that the source strength is updated on-the-fly as the particle travels. This decouples the mesh from any material or geometry considerations. The second is the capability to use a piecewise-linear source approximation, as opposed to a piecewise-constant approximation, with a history-based linear discontinuous tally method. This source tilting technique is designed to reduce the spatial discretization error without the need to refine the mesh.

Numerical experiments were conducted on two k-eigenvalue reactor problems: the 2D C5G7 quarter core problem, followed by the 3D, 2-group, Takeda-1 benchmark problem. Results from the 2D C5G7 benchmark focused on the effects of varying the mesh size with and without source tilting. We started from a very coarse mesh of one mesh cell per pin cell and refined it to a ratio of $8\times8$ mesh cells per pin cell. The ratio of particles per mesh cell was also kept constant for all mesh sizes. As is expected, by refining the mesh size the assembly powers and eigenvalue solutions become more accurate. Furthermore, both assembly powers and eigenvalue solutions were made more accurate by using the history-based source tilting technique often even more so than refining the mesh without source tilting. However, unlike the eigenvalue and MOX assemblies, the $\text{UO}_2$ assemblies did not converge at the expected $O(N^{1-})$ indicating that a finer mesh is necessary in these regions.

Results from the Takeda-1 benchmark showed that using the source tilting technique with a relatively low number of particles helped the region averaged scalar flux and eigenvalue solutions but also introduced significant noise in some regions of the problem. Increasing the number of particles reduced this noise and further improved results. Given this phenomenon, it is easy to imagine that in some problems this noise may hinder rather than help results and there may be cases when a piecewise-constant approximation is more appropriate.  Nonetheless, both the effective source sweep and source tilting techniques were shown to significantly reduce the spatial discretization error. The two techniques complement each other nicely by allowing the user to ``naively'' apply a Cartesian mesh across the domain regardless of the material or geometry used. 

The source tilting technique may be further improved by expanding to include higher order terms through additional moment tallies in the transport sweep. However, the number and complexity of the tallies necessary would grow rapidly and computational performance would be a significant concern. Another potential improvement would be in developing a continuous form of the history-based source tilting. Continuous source tilting techniques have been shown to significantly improve results over discontinuous techniques in IMC methods~\cite{smedley2015asymptotic, shi2020continuous}. Ultimately, to further improve iQMC, attention will need to shift to performance benchmarks, ideally on larger problems like the 3D C5G7. These larger problems will require the use of more advanced domain decomposition techniques but will ultimately provide a more useful comparison to multigroup Monte Carlo or other hybrid transport methods like The Random Ray Method~\cite{Tramm2017}.

\pagebreak
\section*{Acknowledgments}
This work was funded by the Center for Exascale Monte-Carlo Neutron Transport (CEMeNT) a PSAAP-III project funded by the Department of Energy, DE-NA003967, 
and supported by National Science Foundation Grants
DMS-1745654,
and
DMS-1906446.

Special thank you to John Tramm for his consultation in developing k-eigenvalue capabilities for iQMC. We also gratefully acknowledge the computing resources provided by Livermore Computing at Lawrence Livermore National Laboratory. Finally, thanks to the entire MC/DC development team for providing an excellent Monte Carlo test-bed environment.

\bibliographystyle{style/ans_js}                                             
\bibliography{main}


\end{document}